\documentclass[authoryear,12pt,3p,letter]{jowarticle}
\usepackage{graphicx}
\usepackage{amsthm,amsmath}
\usepackage{amssymb}
\usepackage{amsfonts}
\usepackage{natbib}
\usepackage{setspace}
\usepackage[all]{xy}
\usepackage{enumitem}
\usepackage{titlesec}
\usepackage{mathrsfs}
\makeatletter
\renewcommand\section{\@startsection{section}{1}{\z@}{-3.25ex plus -1ex minus -.2ex}{1.5ex plus .2ex}{\normalsize\bf}}
\renewcommand\subsection{\@startsection{subsection}{2}{\z@}{-3.25ex plus -1ex minus -.2ex}{1.5ex plus .2ex}{\normalsize\bf}}
\renewcommand\subsubsection{\@startsection{subsubsection}{3}{\z@}{-3.25ex plus -1ex minus -.2ex}{1.5ex plus .2ex}{\normalsize\bf}}
\makeatother

\newtheorem{innercustomgeneric}{\customgenericname}
\providecommand{\customgenericname}{}
\newcommand{\newcustomtheorem}[2]{%
  \newenvironment{#1}[1]
  {%
   \renewcommand\customgenericname{#2}%
   \renewcommand\theinnercustomgeneric{##1}%
   \innercustomgeneric
  }
  {\endinnercustomgeneric}
}

\newtheorem{thm}{Theorem}

\newtheorem{defn}[thm]{Definition}

\newcustomtheorem{prop*}{Proposition}
\newcustomtheorem{cor*}{Corollary}

\begin{document}
\begin{frontmatter}
\title{Why Not Categorical Equivalence?}
\author{James Owen Weatherall}\ead{weatherj@uci.edu}
\address{Department of Logic and Philosophy of Science\\ University of California, Irvine}
\begin{abstract}In recent years philosophers of science have explored categorical equivalence as a promising criterion for when two (physical) theories are equivalent.  On the one hand, philosophers have presented several examples of theories whose relationships seem to be clarified using these categorical methods.  On the other hand, philosophers and logicians have studied the relationships, particularly in the first order case, between categorical equivalence and other notions of equivalence of theories, including definitional equivalence and generalized definitional (aka Morita) equivalence.  In this article, I will express some skepticism about categorical equivalence as a criterion of physical equivalence, both on technical grounds and conceptual ones.  I will argue that ``category structure'' (alone) likely does not capture the structure of a theory, and discuss some recent work in light of this claim.\end{abstract}
\end{frontmatter}
\doublespacing
\section{Introduction}\label{sec:introduction}

In \citep{WeatherallTE}, I proposed a criterion of equivalence for (physical) theories.\footnote{This proposal was inspired and strongly influenced by \citet{Halvorson}, who had previously argued that category theory would likely be useful for representing theories, particularly for settling questions of (in)equivalence, but who did not make a concrete proposal for how to do this in practice.  Since then, there has been a large literature on this subject in philosophy of physics.  (Of course, mathematicians have long used category theory to explore related issues!)  For a detailed review of the literature on theoretical equivalence in physics, see \citet{WeatherallReview}.}    This criterion states that two theories are equivalent if:
\begin{enumerate}
\item their categories of models are equivalent;\footnote{I do not review basic ideas from category theory in any detail here; for background, see \citet{MacLane}, \citet{vanOosten}, or \citet{Leinster}.} and
\item the functors realizing that equivalence preserve empirical content.
\end{enumerate}
A ``category of models'', here, is a category whose objects are models of a physical theory, and whose arrows are maps that, in a suitable sense, preserve the structure of those models.\footnote{In many physical theories, there are natural candidates for the ``models of the theory''; the arrows require more careful attention, though in practice, ambiguities concerning what one should take as arrows of the category of models reflect real interpretational differences.  For instance, the models of general relativity are relativistic spacetimes, which are four-dimensional manifolds, satisfying certain topological conditions, endowed with a smooth metric of Lorentz signature.  One natural choice of arrows for this category are the isometries, which are diffeomorphisms between spacetime manifolds that preserve the metric.}  What is meant by ``empirical content'', meanwhile, is contextual and difficult to pin down in general; in cases of interest one can make precise the sense in which it is preserved.  I will have much to say about this criterion below, but in the first instance the idea behind it is that two theories are equivalent if (a) their mathematical structures are equivalent, qua mathematics; (b) they are empirically equivalent; and (c) these two equivalences are compatible.  The first of these, (a), is captured by the equivalence of categories; the (b) and (c) are captured by the requirement that it be precisely those functors that realize the categorical equivalence that also preserve empirical content.

As I describe in section \ref{sec:success}, this criterion of equivalence, which (at risk of ambiguity) I will call \emph{categorical equivalence} of physical theories, has been fruitfully employed in a number of cases to clarify the senses in which candidate pairs of theories are equivalent or inequivalent.  I think it has led to correct verdicts in all of these cases, and that at least in some cases it has done so in a way that provides new insight into the theories in question.  But despite these successes, I think there are reasons to be cautious about the adequacy of categorical equivalence as a criterion of theoretical equivalence.\footnote{One might be skeptical that there is any single criterion---that is, any necessary or sufficient conditions---for equivalence of theories; instead, one might think that there are many criteria out there that each capture different senses of equivalence, and that the most fruitful approach is to develop a bestiary of such criteria and to ask, in particular cases, in which senses theories are and are not equivalent.  From this perspective, the worry is that categorical equivalence, as described in the literature, may not adequately capture the sense of equivalence that it is intended to capture---or, perhaps, any interesting sense at all.}  In particular, I will argue, a category of models does not (necessarily) capture the (mathematical) ``structure'' of a theory.  Thus, the (categorical) equivalence of two categories of models does not (necessarily) preserve the structure of a theory.\footnote{Note, here, that the ambiguity alluded to above matters: at issue is whether the categorical equivalence of two categories of models adequately captures the required relationship between the mathematical structures invoked in physical theory; if not, then categorical equivalence, as a criterion of theoretical equivalence that includes an additional condition regarding empirical significance, arguably fails.}  Categorical equivalence is plausibly a necessary condition for equivalence, but it is not sufficient.

The worries I will express are related to, and partially inspired by, considerations previously raised by \citet{Barrett+HalvorsonME} and \citet{Hudetz}.  But the perspective I offer is different.  In particular, I will start with what I take to be the successes of the categorical equivalence approach, and then question whether these successes are achieved for the adverted reasons.

The remainder of the paper will proceed as follows.  I will begin by introducing categorical equivalence in more detail, and then reviewing some of the ways in which categorical equivalence has been studied and applied by philosophers of science.  I will then present some critiques of the approach---including several that I think are important to record, but which I will not pursue (or respond to) here.  Next I will focus in on one particular worry, regarding whether a category can be said to represent the structure of a theory.  I will present what I call the `G' property, which is a property that a category may (or may not) have, and the failure of which in particular examples seems to capture a sense in which those theories do not adequately capture the structure we care about.  I will ultimately argue that the `G' property is neither necessary nor sufficient for a category to capture this structure, but I will suggest that it points in a fruitful direction.  In the final section, I will describe three attitudes one might adopt given the arguments I present, and argue that they suggest different---though not necessarily mutually exclusive---research programs that one might pursue.

\section{Categorically Equivalent Theories}\label{sec:success}

Categorical equivalence has been explored by philosophers from two directions.\footnote{\label{fn:IE} A third direction---pursued, for instance, by \citet{NguyenTE} and by \citet{Butterfield}---has been more critical.  These authors argue that there are further considerations, related to the interpretation of theories, that are necessary to establish equivalence.  (See also \citet{Coffey} and \citet{Sklar}.)  In a sense, this is uncontroversial: after all, categorical equivalence, as described above, requires theories to be empirically equivalent, in a way that is compatible with their categorical equivalence, and empirical equivalence depends on interpretation.  But there is still a matter of controversy over whether there are \emph{further} senses in which interpretation should matter.  I set this cluster of issues aside in what follows, as my goal is to raise a different set of concerns about categorical equivalence.}  One, largely theoretical approach---pursued, for instance, by \citet{Barrett+HalvorsonME} and \citet{Hudetz}---has focused on properties of categories of models of theories in first or higher order logic.  This approach has established that, in the first order case, categorical equivalence is strictly weaker than other compelling notions of equivalence, such as definitional equivalence and so-called ``Morita'', or generalized definitional, equivalence.\footnote{Generalized definitional equivalence of physical theories has been studied extensively by \citet{Andreka+NemetiNotes}; see also \citet{Andreka+NemetiComparing} and \citet{Lefever+Szekely}.}  Insofar as these other notions of equivalence, which seem natural within a logical context, are well-motivated, one might take this result alone to show that categorical equivalence is too weak.

But this conclusion may be too fast, because it is not clear ``how much'' weaker categorical equivalence is.  In particular, \citet{Barrett+HalvorsonME} conjecture that categorical equivalence is equivalent to Morita equivalence (in the sense of yielding the same verdicts) for certain classes of first order theories,\footnote{Specifically, they conjecture that categorical equivalence implies Morita equivalence for theories with finite signatures.} and they suggest that since categorical equivalence is much easier to apply in cases of real physical theories, where one often does not have a first order formulation.  For just this reason, we do not have examples of (physical) theories that are categorically equivalent but inequivalent in some stronger, logical sense, of a sort that we might want if we were to evaluate what is missing from categorical equivalence.

This remark brings us to the second direction from which categorical equivalence has been studied by philosophers.  What we \emph{do} have is a growing handful of physical theories that are either categorically equivalent, or else which fail to be categorically equivalent in ways that help us to better understand their relationship.\footnote{See \citet{WeatherallLandry} for a more detailed discussion of the ways in which categorical equivalence has been applied to better understand classical field theories.}  In particular, categorical equivalence has been used to argue for the equivalence of several pairs of theories: electromagnetism formulated in terms of vector potentials and in terms of electromagnetic fields are equivalent if and only if one takes vector potentials related by a gauge transformation to be isomorphic \citep{WeatherallTE,WeatherallUG}; likewise, Newtonian gravitation and geometrized Newtonian gravitation (Newton-Cartan theory) are equivalent if and only if one takes gravitational potentials related by a certain class of transformations to be isomorphic \citep{WeatherallTE}; Einstein algebras and general relativity are equivalent \citep{Rosenstock+etal}; Lagrangian mechanics and Hamiltonian mechanics are equivalent on one way of conceiving of each theory, but one can motivate other ways of thinking of these theories on which they are not equivalent \citet{BarrettCM2}; and likewise there are various formulations of Yang-Mills theory that are equivalent and inequivalent \citep{WeatherallUG, Rosenstock+Weatherall, Nguyen+etal, Bradley+etal}.

The details of these examples do not matter for what follows.  The point is that the application of categorical equivalence to these cases of antecedent interest has been fruitful.  For one, I believe that categorical equivalence has given the right verdict in all of these cases: the theories that are categorically equivalent \emph{are} equivalent, in the most salient senses; and the theories that are not categorically equivalent are \emph{not} equivalent---and more, the ways in which they fail to be categorically equivalent have provided insight into the theories in question.  In particular, in the cases where one establishes \emph{inequivalence}, one can still identify functors that relate the relevant theories' categories of models, and which do so in a way that preserves empirical equivalence.  Studying the properties of such functors can reveal information about the theories.  For instance, if one does not take vector potentials related by gauge transformations to be isomorphic to one another, then there is a precise sense in which vector potentials have more structure than electromagnetic fields.

Of course, the idea that vector potentials have ``excess structure'', as compared with electromagnetic fields, is not surprising; indeed, it is best seen as a litmus test for whether the approach is plausible.\footnote{Though whether the sense of excess structure just alluded to is the right one to consider is disputed---see \citet{Nguyen+etal}, and \citet{Bradley+etal} for a response.}  But in other cases, the results have been more novel.  For instance, there is a certain theme in the philosophy of physics literature according to which to be a realist about the standard formulation of general relativity is to endorse some form of substantivalism.\footnote{This idea is present throughout much of the literature on the hole argument \citep{Earman+Norton,NortonSEP}; see \citet{BrighouseNew} for a particularly clear statement of the view, and \citet{WeatherallStein} for a discussion of its origins.}  The reason is that general relativity is standard formulated as a theory of fields on a four-dimensional manifold of spacetime ``points''; some authors have concluded that this means general relativity posits spacetime as an independent entity, ontologically prior to matter.  In contrast, \citet{Earman1986, EarmanWEST} has suggested that Einstein algebras are an appropriate formal setting for a form of \emph{relationism} about spacetime in general relativity.  Briefly, an Einstein algebra is an algebraic structure whose elements represent possible configurations of fields \citep{GerochEA}.  Beginning with such a structure, one can proceed to define the structures one uses in general relativity: a metric, a derivative operator, and so on; and one can express Einstein's equation.  But one does so without ever introducing a manifold, and so one might think that this is a theory that does not posit, as a primitive entity, a spacetime manifold.  Instead, one works only with possible configurations of matter, relations between those configurations, and structures one can define on the algebra of such configurations.

But, drawing on results due to \citet{Nestruev}, \citet{Rosenstock+etal} show that these two theories are categorically equivalent.\footnote{There were hints in the philosophical literature that something like this should hold.  In the first instance, \citet{Earman1986,EarmanWEST} suggested that Einstein algebras might resolve the apparent indeterminism in general relativity revealed by the hole argument \citep{Earman+Norton}.  But as \citet{Ryno} argued, in related cases in algebraic topology the sorts of maps that are used to run the hole argument also arise between algebras of functions, suggesting that it is hopeless to expect the hole argument to go away if one moves to Einstein algebras.  (See also \citet{Bain}.)  This relationship between the maps between Einstein algebras and relativistic spacetimes is at the heart of the equivalence result.  For a different view of the hole argument, compatible with the attitudes adopted in the present paper, see \citet{WeatherallHoleArg}.}  This result shows more clearly how one can think of a relativistic spacetime as a means of encoding, or representing, nothing more or less than the possible ways in which matter may be (spatio-temporally) configured and the relations between those possible configurations, precisely as Einstein algebras do.  This, I think, substantially clarifies the structure represented by a relativistic spacetime.

In other cases, exploring functors between physical theories has provided new insight into otherwise murky philosophical disputes.  For instance, in a series of recent papers addressing the relationship between Lagrangian mechanics and Hamiltonian mechanics, \citet{North} has argued that these theories are inequivalent and that Hamiltonian mechanics has ``less structure'' than Lagrangian mechanics; \citet{Curiel} has argued that they are inequivalent because Langrangian mechanics has ``less structure'' than Hamiltonian mechanics; and \citet{BarrettCM1} has argued that the structures North and Curiel focus on are actually incomparable, in a certain precise sense.  More recently, \citet{BarrettCM2} has substantially clarified the situation, by showing that there is a precise sense in which Lagrangian and Hamiltonian mechanics are (categorically) equivalent, if one takes Lagrangian mechanics to be a theory in which one defines a Lagrangian on the tangent bundle of a configuration space, and Hamiltonian mechanics to be a theory in which one defines a Hamiltonian on the cotangent bundle of a configuration space.

But Barrett also argues that there are other, well-motivated ways of understanding these theories---one of which he identifies with North, and one of which he identifies with Curiel.  He then shows that, if one follows North's proposal, there \emph{is} a sense, after all, in which the theories are inequivalent.  And likewise, if one follows Curiel's proposal, the theories are also inequivalent.  And so one sees that the dispute comes down to what one should mean by ``Hamiltonian mechanics'' and ``Lagrangian mechanics''.  Of course, these technical results do not resolve this background issue.  But they help to isolate precisely where the disagreements lie.

I take these results and the arguments based on them to be good reason to take categorical equivalence seriously.  More, since I think that categorical equivalence has yielded the correct verdicts in each of these cases, and done so in a precise way, I think that any critique of categorical equivalence needs to account for \emph{why} using categorical equivalence has been fruitful in these ways.  But I also think that categorical equivalence is probably not the correct criterion of equivalence, for reasons I will elaborate in the next several sections.

\section{Interlude: Concerns I will not Pursue}

The present section lies largely outside the main thread of the paper and can be freely skipped.  But since the goal of the paper is to question the adequacy of categorical equivalence, I will now note a number of concerns about the criterion that I think are largely distinct from the concern I focus on in the following sections.  I think these, too, are issues that will need to be addressed in any successful future development of the proposal.\footnote{I do not take the list here to be exhaustive!  For instance, I do not (otherwise) raise any concerns about using groupoids---categories where every arrow is invertible---as opposed to categories with richer arrow structure, to represent theories.  I also note that \citet{Barth} has also criticized categorical equivalence---or rather, what he has called the solution-category approach---along lines that are related to some of the points below, though I do not reproduce all of his concerns here and I think some of my concerns are different.  Still, my thinking has certainly been influenced by his.}  I will not pursue these in any detail; I do not think they are dispositive, but I will not try to refute them.

The first concern has to do with the heterogeneity of the examples noted in the previous section.  In particular, these examples involve subtly different conceptions of what counts as a ``model'' of a physical theory---that is, what one takes to be the objects in a ``category of models''.  This is so even though in virtually all the cases discussed above, the mathematical structures under consideration are of a similar character: they tend to be smooth manifolds endowed with some further structure, often represented by fields.  But the manifolds in the different cases have different representational significance.  This would not be a problem were one concerned only with equivalence of theories in a purely logical or mathematical context, where one might take a general view of models as structures that realize some axioms.  But in the context of scientific theories, it becomes troubling.  This is because it suggests a lack of clarity about what sorts of structures are the relevant ones to count as models for a scientific ``theory'' in the wild.  Given a theory, as described in a textbook or review article, say, how is one to decide how to identify the relevant models to construct the required category?

For instance, in the examples of (a) general relativity and Einstein algebras, (b) Newtonian gravitation and geometrized Newtonian gravitation, and (c) vector potential and electromagnetic field formulations of electromagnetic theory, the models of the theories in question are structures representing the complete history of the universe.  In all of these cases, except for Einstein algebras, one considers a manifold of (all) events in space and time on which various fields are defined; in the Einstein algebra case there is no manifold, but one has algebras of global, in space and time, field configurations.

But in other cases this is not what one does.  For instance, in the example of Lagrangian and Hamiltonian mechanics---which should not be dismissed as an outlier, since as I argued above, it is an example in which we have learned something about the relevant theories---the models of the theories are taken to be structures representing all possible instantaneous states of a system, along with a particular dynamics governing the evolution of that system: in the case of Lagrangian mechanics, this is a configuration space along with a Lagrangian function; whereas in Hamiltonian mechanics, it is a phase space along with a Hamiltonian function.  A complete history of a system---the models of the first class of examples---would be a trajectory through one of these spaces, rather than the space itself; conversely, in the first class of examples one does not consider different possible dynamics, but only different solutions for a single dynamics.

Other examples are different, still.  Consider the case of Yang-Mills theory, for instance.  On a fiber bundle formulation of the theory, the models are taken to be principal bundles over a manifold of events, with a metric on the base manifold and a principal connection on the principal bundle.  And so in a sense there are two manifolds: one representing the global evolution of the universe in space and time, and another carrying information about possible configurations of matter at each point, but without picking out any particular such configuration.\footnote{How different really is this situation from the structure one considers in general relativity?  It would take us too far afield to evaluate this question in detail, though \citet{WeatherallYM} argues that the structures are much more similar in character than they initially appear.}  One might also consider still other candidates that would be natural in theories that have not been studied with these methods.  For instance, one might take a candidate ``model'' of quantum theory to be a Hilbert space along with a $*$ algebra of operators acting on that Hilbert space; in this case, the model would represent a state space along with a privileged set of observables.

But this last observation raises another issue, which also speaks to the lack of clarity regarding what is meant by ``models'' here.  Despite the heterogenity just noted, there is also a striking homogeneity in the examples discussed in the last section---and those studied with these methods.  In particular, they are all examples from classical physics, and except for Hamiltonian and Lagrangian mechanics, they are all ``classical field theories''.  But this is surely a small subset of the physical theories that may be equivalent to others!  Where are the examples from quantum mechanics, quantum field theory, statistical physics, and elsewhere?  The question is particularly striking given that some of the classic examples of ``(in)equivalent theories in the wild'' are from the history of quantum theory, such as the famous example of Schr\"odinger's wave mechanics and Heisenberg's matrix mechanics.  Can categorical equivalence capture the sense in which these theories are equivalent?  Arguably the reason no one has done so yet turns on precisely the ambiguity noted already concerning what should count as a model of a theory.  It is not perfectly clear what the category of models of wave mechanics or matrix mechanics ought to be.

Yet another cause for concern, largely orthogonal to the issues already raised, is that the sense in which philosophers use category theory to represent physical theories for the purposes of establishing equivalence and inequivalence seems pretty different from the ``native'' applications of categories in physics, such as in the contexts of locally covariant (quantum) field theory \citep{Brunetti+etal} or (higher) gauge theory \citep{Baez+Schreiber}.  In these cases, one tends to use category theory to \emph{express} the models of the theory, not (in the first instance) to relate those models.  So it seems that category theory is entering at a different level of analysis.\footnote{On the other hand, once one has used categories to build models of a theory, it is often natural to construct categories of such models.  But when one does so for theories whose constructions are ``native'' to category theory one tends to get much richer structures.}

The final two concerns are more general in character, and have less to do with the particular examples that have been studied.  First, recall that the criterion of equivalence given above requires not only categorical equivalence, but equivalence given by functors that ``preserve empirical content''.  The idea is that two theories must be, at least, empirically equivalent if they are to be ``theoretically'' equivalent.  But the notions of ``empirical content'' and ``empirical equivalence'' are not clear, and it is hard to see how they could be made precise.\footnote{An anonymous referee notes that attempts to make ``empirical equivalence'' precise exist in the literature---as, for instance, in \citet{VanFraassenSI}.  Fair enough.  But as van Fraassen himself would surely acknowledge, especially given his more recent work, how our scientific theories come to represent worldly situations involves a great deal of interpretation, intention, practice, and context \citep{vanFraassenSR}.  One can develop formal tools for expressing these intentions, etc., at which point precise standards of equivalence may be employed.  But this hardly yields a formal test of empirical equivalence analogous to the way in which definitional equivalence, say, is a formal test of equivalence first order theories.  The worry, which I articulate presently, is that the vast bulk of the work of establishing equivalence occurs when one tries to take the messy details of how a scientific theory is used to represent the world and distill them into the formal apparatus---and not when one checks the properties of a certain functor.}  Indeed, one might worry that it is a prior, largely unanalyzed, notion of which theories are empirically equivalent that is doing the work in the examples discussed above, and that the category theoretic analysis adds little to this.\footnote{This concern is implicitly raised by \citet{NortonUnderdetermination}, who argues that whenever two theories are empirically equivalent, it follows that they must share so much common structure that they are best conceived as notational variants.  Though see \citet{Bradley} for a reply.}  (I return to this point in section \ref{sec:roots}.)

Worse, one might worry that one cannot establish ``empirical equivalence'' once and for all: one can, at best, establish it for some class of possible interactions or possible measurements.  Indeed, in the case of the two formulations of electromagnetism discussed in the previous section, it is natural to take the empirical predictions of each theory to be captured by the measurements one could make with (classical) charged matter.  From this perspective, the two theories are empirically equivalent, insofar as they predict the same motions of charged matter.  But if one includes interactions with matter represented quantum mechanically, this equivalence breaks down, as shown by the Aharanov-Bohm effect, which is a measurable effect on the behavior of a particle propagating in a region of vanishing electromagnetic field but non-vanishing vector potential.\footnote{The significance of this example for our understanding of classical electromagnetic theory is discussed by \citet{BelotUE}.}  This suggests that any equivalence one establishes is, at best, provisional, since future developments elsewhere in physics could lead to experiments that could discriminate between allegedly ``equivalent'' theories.  At the other extreme, one could also imagine forcing equivalence by artificially restricting the possible measurements under consideration.

Finally, and most generally of all: one might be skeptical about the idea of having a single, once and for all, characterization of a physical theory---be it as a category, a set of sentences in some language, or even a textbook account.\footnote{Note that this concern is a bit different from that raised by the authors noted in footnote \ref{fn:IE}, because it is not specifically about \emph{formal} representations of a theory.  Rather, the concern is that theories are not to be pinned down at all, whether formally or not.  Indeed, one might take the remarks here to be a kind of skepticism about any account of the ``structure of theories'' or even the ``semantics of theories''.  For an interesting discussion of these issues, see \citep{FormicaFriend}, who argue that physical theories should be associated with a network of different formal axiomatic theories, rather than a single theory.  Thank you to Hajnal Andr\'eka and Istvan N\'emeti for drawing my attention to this work.}  The idea is that physical theories are messy affairs including all sorts of arguments, intuitions, biases, interpretations, intentions, numerical methods, and so on.  Worse, they are dynamic: methods develop and improve, arguments once found persuasive are rebutted, strongly held intuitions are abandoned, and so on.  General relativity as conceived today---insofar as there is an univocal thing that goes by that name---is radically different from general relativity as understood by Einstein and his collaborators.  But given this ever-changing richness, in what sense could one ever hope to identify a single category associated with a physical theory, much less establish that, because of some relation that a category stands in to another category, that two theories are ``the same''?

\section{Category Structure and Ideology}

As I indicated above, I will not pursue the concerns described in the previous section.  Instead, I will focus on a worry that I think is more central to the program.  Put briefly: a ``category (of models)'' does not (necessarily) capture the ``structure'' of a theory.  Or in other words, there is more to the mathematical structure of a theory than just the ``category structure'' of its models.

To say what I mean by this, I first need to say what I mean by ``category structure''.  Here I adopt a certain ideological posture, which is that the structure of a mathematical ``gadget'' of a given kind is to be sharply distinguished from the procedure by which you came to construct that object.\footnote{The term ``gadget'', here, comes from John Baez, who introduced it (in conversation) because ``object'' has a technical meaning in category theory and ``structure'' seems to carry too much baggage; basically, a mathematical ``gadget'' is any sort of thing that mathematicians define and study.  I adopt and discuss this ideology in \citet{WeatherallHoleArg}; I think it is particularly clearly expressed in \citet{Burgess}, though he does not go on to emphasize the relationship to ``structure-preserving maps'' that I introduce presently.  Although the basic point is very closely related to famous arguments by \citet{Benacerraf}, it is important to distinguish the ideology about mathematical gadgets that I am adopting from what is sometimes called ``mathematical structuralism''.  Structuralism, as I understand it, is a view about the ontology of mathematical objects.  I do not mean to make any particular claims about ontology here, and I take the claims I make here to be compatible with many different philosophies of mathematics.}  For instance, one may think of the group $Z_2$ as the quotient group of the integers by even integers, as the symmetry group of a set with two elements, or as the sphere group $S^0$, i.e., the group of real numbers of unit length.  One might also think of it as some set of ordered sets, the first of which is a domain, and the rest of which characterize multiplication, the identity element, and so on.  All of these different ways of constructing $Z_2$, however, are the same in one respect: they all have, or instantiate, the same group structure.  They are the same, as groups.  I take this to mean that, insofar as one wishes to use (just) the group $Z_2$ for some representational purpose (as opposed to using some other mathematical gadget, perhaps with more or different structure), the construction procedure cannot matter to the success of the representation or the validity of inferences drawn from it.  In other words, one is using $Z_2$ for some representational purposes only insofar as one does not use features of an instantiation of $Z_2$ that are not shared by all instantiations.

But how are we to isolate, from a particular realization of some gadget---say, $Z_2$---precisely what structure is intended?  The key is to study the maps that preserve that structure.  Of course, we do not get such maps for free.  But I take it that an essential part of mathematical practice is to define, whenever a new mathematical structure is proposed, a class of mappings that preserve the intended structure---that is, generally, but not always, the isomorphisms of the structure.  It is these mappings that capture the sense in which the different realizations of $Z_2$ are ``the same'': they are all related by group isomorphisms.  And by looking at what these mappings preserve, one can say what the structures are.  For instance, group homomorphisms are mappings between groups that preserve group multiplication; one can infer from this that groups are collections of elements that are distinguished from one another (only) by their multiplicative relations with other group elements.  In the case of $Z_2$, this means: $Z_2$ has exactly two distinct elements, one of which is the identity and the other of which, when multiplied by itself, yields the identity.  What these elements are, or which element realizes which properties, is not part of the structure, as this is not preserved by group isomorphism.

From this perspective, we can now return to category structure.  Here the maps that preserve (all) categorical structure are precisely the categorical equivalences.\footnote{Why not the categorical isomorphisms---that is, the invertible functors?  One certainly could take these to be the relevant standard of equivalence, but this is rarely done in category theory.  One reason is that categorical isomorphisms preserve ``too much'', in the sense that the intended structure of a category does not include a determinate number of objects (only a determinate number of non-isomorphic objects).  Another, related, reason is that it is often natural to think of functors as themselves having structure-preserving maps, known as natural isomorphisms, between them.  Categorical equivalences are functors that have ``almost'' inverses---that is, inverses up to natural isomorphism.  In this sense, categorical equivalence might be understood as isomorphic up to isomorphism.}  Hence, by reflecting on what is preserved by categorical equivalence, we learn what category structure is.  In analogy to groups, we find: a category is a collection of objects distinguished (only, and only up to isomorphism) by their arrow-algebraic relations with other objects.  In other words, we find that the arrows carry all the information; objects are essentially placeholders.\footnote{These remarks are not meant to be surprising, particularly to anyone who is accustomed to working with categories.  In fact, category theorists often emphasize that it is the arrows that do all the work, and there are even single-sort axiomatizations of category theory in which only arrows appear.}

In particular, this means that the ``internal structure'' of objects is not (automatically) preserved under categorical equivalence.  Indeed, there is a classic concern along these lines, expressed by \citet{Hudetz} and others, that categorical equivalence may ``trivialize'' in some cases.  Consider some well-understood, concrete category---say, the category whose objects are groups and whose arrows are homomorphisms.\footnote{I am not worrying, here, about size considerations.  But for someone who is worried about whether the category of groups is well-defined, we can consider all groups of cardinality less than some inaccessible cardinal, $\kappa$.}  Now consider a category whose objects are giraffes, but whose arrows are chosen so that there exists an equivalence between the category of groups and the category of giraffes (with specially chosen arrows).  By construction, these are equivalent.  But groups and giraffes are not the same!  Does this not immediately imply that categorical equivalence is too weak a notion of equivalence?

Perhaps.  Indeed, from a certain point of view, this claim should not be surprising, for reasons discussed at the beginning of section \ref{sec:success}: even in the first order case, categorical equivalence is weaker than Morita equivalence, and so if one thought that Morita equivalence was the ``right'' notion of equivalence for first order theories, one should conclude that categories do not capture the structure of theories.  One response to this sitation, proposed by \citet{Hudetz}, is to strengthen the notion of ``categorical equivalence'', by adding further constraints on the functors that realize the equivalence.  In particular, Hudetz suggests that these functors should be \emph{definable}, in the sense that if the functor takes an object of one category to an object of another, then the object of the codomain category (or an isomorph thereof) should be definable in terms of the object of the domain category.\footnote{Hudetz makes this condition precise, but for present purposes an informal description suffices.}  I will return to this proposal in Sec. \ref{sec:roots}.  But for now, let me simply remark that this proposal is inconsistent with the ideology described above, at least if we understand theories to be represented by categories, precisely because it invokes the internal structure of objects.

But there is another possible response to this concern, which is to deny that the trivialization concern is real.  The idea is that, at least in some cases, category structure \emph{can} represent the structure of a theory, precisely because the internal structure of the objects of the category is suitably reflected, or encoded, in the arrows of the category.  If this is right, then the claim that the objects of a category are ``giraffes'' is immaterial---just as claiming that some realization of the group $Z_2$ happens to have, as elements, giraffes, who have a certain multiplication relation defined on them.\footnote{There is a connection here to classic work by \citet{Makkai}, on a duality between syntax and semantic in first order theories; see also \citet{Awodey+Forssell,Lurie}.  At least in some cases, one can reconstruct a theory, uniquely up to a suitable notion of equivalence, from its category of models.  But the relationship between such results and the categories encountered in the philosophy of physics literature is not clear.}  To see how this goes, consider an example: the category of sets, \textbf{Set}, whose objects are sets and whose arrows are functions.  (Of course, by the above, that the objects are sets is immaterial---I am simply giving a construction.\footnote{In fact, the category of sets can be defined ``directly'', as the category satisfying certain axioms, as opposed to by beginning with a prior definition of sets and functions.  See \citet{Lawvere}}.)  It turns out that in this case, using only arrow constructions, we can reason about sets in detail.  For example, there is an object in this category, unique up to isomorphism, that has the property that there exists a unique arrow from any other object to this object.  Call this object $\mathbf{1}$.  (It happens to be a set with one element.)  Then arrows \emph{from} $\mathbf{1}$ may be thought of as elements of their codomains.  Similarly, we may think of monomorphisms (i.e., injective functions) as subsets, constructions known as coproducts are disjoint unions, and so on.  All of this is true irrespective of what we happen to say the internal structures of the objects are like.

This example seems to show that the trivialization concern is chimerical: much more is (or can be) encoded in the arrows of a category than is immediately apparent.  But is this always the case?  As I argue in the next section the answer appears to be ``no''---including for cases of interest for the categorical equivalence program in philosophy of science.

\section{The `G' property}\label{sec:Geroch}

In the previous section, I argued that in the category \textbf{Set}, the internal structure is ``externalized'', i.e., that the internal structure of sets is reflected in the arrows of the category. One way of understanding how this works is to note that objects of the category are uniquely distinguished, up to isomorphism (bijection), by their positions in the graph of arrows.  For instance, two sets are isomorphic if and only if they have isomorphic automorphism groups.  This feature of \textbf{Set} suggests a proposal.  Perhaps one should say that a theory is captured by category structure only if the arrows of that category can distinguish the objects, up to isomorphism.

More precisely, we define the following property that a category may have.\footnote{I call this the `G' property because it was proposed by Bob Geroch during a conversation with Hans Halvorson at a meeting in Pittsburgh in April, 2013.  Essentially the same condition was also discussed, apparently independently, by \citet{Dewar+Eva}, though their motivation for considering the condition was different: they suggested that violating it would indicate that a theory has ``excess structure''.  I do not engage further with their proposal here.  It is also considered by mathematicians, under a different name: the `G' property is precisely the condition that the automorphism class group of a category be trivial \citep[cf. ][Problem 1.B]{Freyd}.  (I am grateful to Hans Halvorson for bringing Freyd's work to my attention back in 2013.)}
\begin{defn}A category $C$ has the \emph{`G' property} if every full, faithful, and essentially surjective functor $F:C\rightarrow C$ is naturally isomorphic to $1_C$.
\end{defn}
In other words, a category has the `G' property if every ``autoequivalence'', that is, every symmetry of the category, in the sense of an equivalence of the category with itself, is naturally isomorphic to the identity.\footnote{Here we make use of the fact that every full, faithful, and essentially surjective functor as an almost-inverse, i.e., is an equivalence of categories.  For definitions of full, faithful, and essentially surjective, see \citet{Leinster}.}  Spelling this out, it means that any way of mapping objects of a category to objects of a category that preserves the network of arrows will necessarily take objects to isomorphic objects.  In this sense, then, the condition captures the idea that objects are distinguished, up to isomorphism, by their place in the network of arrows.

The category \textbf{Set} has the `G' property.  So do other ``concrete'' categories that one often encounters, such as the category \textbf{Group} of groups and group homomorphisms \citep[p. 31]{Freyd}. But it is also easy to identify simple categories for which it fails: consider, for instance, a category with two objects and two arrows (the identity on each object).  There is an autoequivalence that swaps the ojects, but the objects are not isomorphic (and indeed, there are no arrows between them).  Still, such examples are contrived, and it is hard to see what internal structure is being captured (or not) in such a case.  And so one might ask: do categories that we might naturally associate with, say, physical theories always have the `G' property?

The answer is ``no''.  Consider, for instance, general relativity.  We might define a category of models of \textbf{GR} as follows: it is a category whose objects are relativistic spacetimes---that is, smooth, Hausdorff, paracompact four dimensional manifolds with smooth Lorentz-signature metrics---and whose arrows are isometries---that is, diffeomorphisms that preserve the metric.  (This is the category associated with \textbf{GR} by, for instance, \citet{Rosenstock+etal}.\footnote{Below I discuss concerns about whether this category has the ``right'' arrows.  But one might also object that it is not clear what \emph{objects} the category should have, on the grounds that it is not clear if we should limit attention to spacetimes that are connected, maximal, etc. \citep{ManchakAN}.  I set this issue aside here, but note that the two questions may interact in interesting ways.})  This category does not have the `G' property.  The reason is that many relativistic spacetimes have no non-trivial symmetries at all---and since the only arrows of the category are isomorphisms, these objects are not distinguished from one another.  Choose any two, non-isometric, spacetimes, each with only one automorphism, and consider a functor from \textbf{GR} to itself that takes the first spacetime (and any spacetime isometric to it) to the second; takes the second (and any spacetime isometric to it) to the first; and acts as the identity on everything else.  (There is only one action on arrows that makes this mapping into a functor.)  The functor so described is an equivalence---but by construction it is not naturally isomorphic to the identity functor.\footnote{One might worry that this argument depends essentially on \textbf{GR} being a \emph{groupoid}---i.e., that it has no non-invertible maps.  This means that there is no information about which spacetimes might be, for instance, embeddable in one another.  Perhaps by adding more arrows, such as isometric embeddings, one could produce a category that has the `G' property.  But I doubt it, because one could then consider spacetimes that were, roughly speaking, asymmetric at all scales.  This would generate more complicated structures, but still no automorphisms.  I suspect that a similar functor could be generated under these circumstances, though I do not claim that this is a proof.}

So the `G' property does not hold of all categories of interest---and, conversely, the category \textbf{GR} does not have the resources to distinguish non-isometric spacetimes.  (Indeed, reflection on the argument just given suggests that \textbf{GR} encodes very little about the generic spacetimes.)  This seems to me to be a serious problem for the categorical equivalence program.  The reason is that the equivalence of categories of models is supposed to capture the sense in which two theories are, mathematically, equivalent.  But there is far more to even the mathematical structure of relativistic spacetimes than is encoded in the category \textbf{GR}.

Still, the situation is not perfectly clear.  There are reasons to think that the `G' property is neither necessary nor sufficient for a category to capture the ``structure'' of a theory.  This property may seem to capture what makes \textbf{Set} seem suitably externalized, but it is not quite what we want.  \textbf{GR} may well be deficient, but the `G' property does not capture why.

To see that the `G' property is not sufficient, consider the following two theories, with associated categories.  First, we have the theory ``Directions''.  Directions says ``the cardinal directions form a two dimensional vector space (over the reals), with `north' and `east' physically distinguished''.  Its category of models, \textbf{Di}, has, as objects, two dimensional vector spaces with (preferred) ordered basis, and as arrows, linear bijections that preserve that (ordered) basis.  Now consider the theory ``Baubles''.  Baubles says ``there are two shiny things, one of which is red and the other of which is blue''.  Its category of models, \textbf{Bau}, has, as objects, ordered pairs (of distinct elements), and as arrows, bijections that preserve order.

One can easily see that both \textbf{Bau} and \textbf{Di} have the `G' property.  This is because, in both cases, all models of the theories are (uniquely) isomorphic to one another, which means that any autoequivalence of the categories of models will necessarily take objects to isomorphic objects.  But it is hard to see how either captures the structure of their respective theories.  Indeed, \textbf{Di} and \textbf{Bau} are equivalent, despite the models having very different internal structures: the objects of \textbf{Di} are two dimensional vector spaces, which have infinitely many elements; whereas the objects of \textbf{Bau} have only two elements.  But the arrows of the categories do not reflect this.\footnote{This sort of situation arises often when one has highly structured (or highly asymmetric) objects in a category.  There are very few maps available that preserve all of the relevant structure.}  So the `G' property does not seem to be sufficient for a category to have suitably captured the internal structure of its objects.

The \textbf{Bau}-\textbf{Di} example may seem contrived---and if so, one might think that there is some other property that it fails to have but which, in conjunction with the `G' property, would capture the desired features.\footnote{There is another response available to the \textbf{Bau}-\textbf{Di} example, which is to say: in fact, the internal structure of the objects in these categories is not so different after all.  This response is motivated by the idea that (adopting the terminology of \citet{Winnie}) the objects of these categories are ``co-determinate'', in the sense that any two dimensional vector space, with ordered basis, is determined ``freely'' by that basis, i.e., by an ordered pair; and every two dimensional vector space with ordered basis determines, in particular, an ordered pair (consisting of the basis elements).  So, perhaps, once we choose an ordered basis for a vector space, the entire vector space structure should be seen as ``determined by'' (or, roughly, definable from) that basis.  From this perspective, that property `G' holds of these categories is not a problem for property `G'.  But alas, this response is too fast.  The reason is that these categories are \emph{too} rigid, and one can easily come up with other categories, equivalent to both, for which this ``co-determination'' relationship does not seem to hold.  Consider, for instance, the category whose objects are sets with one elements and whose arrows are functions preserving that element.  (Or: the category $\mathbf{1}$, with a single object and a single arrow.)  This category is equivalent to both \textbf{Bau} and \textbf{Di}!  And yet it is hard to see how a set with \emph{one} element could determine a two dimensional vector space in any interesting sense (since the free vector space on one element is one dimensional).  I am grateful to Thomas Barrett for pushing me on this point.}  On this view, one might expect the `G' property to be necessary, but not sufficient.  But there are reasons to think this fails as well, as can be seen from the following, non-contrived example.

Consider the category \textbf{Ring} whose objects are rings and whose arrows are ring homomorphisms.  Rings, recall, are Abelian groups endowed with a second operation---multiplication, as opposed to the group operation, which is called ``addition'' in this context---that is associative and distributive over addition.\footnote{Rings are also generally taken to have a multiplicative identity.}  In general, ring multiplication is not commutative, so that the order of multiplication matters.  It turns out, though, that although the order of multiplication matters, there is a certain sense in which the order is nonetheless a matter of convention.  To make this observation precise, note that given any ring $R$, one can always construct an \emph{opposite ring}, $R^{op}$, which has precisely the same elements, but whose multiplication operation is such that for any $A,B\in R$, $A\times_{R} B = B\times_{R^{op}} A$.  Thus, we see that there is a certain intuitive sense in which the ``same'' multiplication relations may be captured using $\times_R$ or $\times_{R^{op}}$---namely, by reading left to right versus right to left.

One is tempted to say that the ring $R$ and the ring $R^{op}$ have ``the same'' ring structure.  After all, they have the same elements, and, in a sense, the same multiplicative relations, simply expressed in a different way.  But in fact, a ring and its opposite are not, in general, isomorphic, and so whatever the intuitive sense in which rings and their opposites ``have the same ring structure'' may be, it is not captured by isomorphism.\footnote{In fact, although many examples of rings that are not isomorphic to their opposites are known, they are not exactly trivial to state.  See \citet[\S 2.8]{JacobsonAlgebra} or \citet[\S 1]{Lam}.  I believe it was Hans Halvorson who first brought this example to my attention.  Observe that groups, too, have opposites, defined in a similar way, but in general groups \emph{are} isomorphic to their opposites, where the isomorphism takes group elements to their inverses.}

Instead, this relationship is captured by an autoequivalence of the category \textbf{Ring}.  The transformation that takes rings to their opposites, and acts in the obvious way on arrows, determines a functor $Op:$\textbf{Ring}$\rightarrow $\textbf{Ring} that takes rings to their opposite rings.  This functor is full, faithful, and essentially surjective, and thus it is an autoequivalence (actually, it is an automorphism) of the category.  But since rings are not isomorphic to their opposites, it immediately follows that the functor $Op$ is not naturally isomorphic to the identity, because it takes objects to objects that are not isomorphic.  Thus, the category \textbf{Ring} does not have the `G' property.

One response to this example would be to concede that the category \textbf{Ring} does not adequately capture the structure of rings.  But I think this is too fast.  As I suggested above, the functor $Op$ seems to take rings to other rings that are, in some sense, ``the same'', but where that sense of sameness is not captured by isomorphism.  In other words, one might think of this autoequivalence as reflecting a real ``symmetry'' of the theory of rings.\footnote{Some readers might be tempted, in light of this, to revise the notion of ``isomorphism'' associated with rings, so that all rings that are suitably ``the same'' are isomorphic.  But this strikes me as a disastrous proposal.  A ring homomorphism that could not distinguish left multiplication from right multiplication would wash out the structure of non-commutative rings!}  Far from failing to capture the structure of rings with the arrows of the category, the $Op$ functor shows that the network of arrows captures a non-trivial sense in which non-isomorphic rings can nonetheless be the same.  Of course, this is not a proof of anything, because I have not provided a definitive argument that \textbf{Ring} \emph{does} capture the structure of the theory of rings.  The argument is merely suggestive that the `G' property is not necessary.  But whatever else is the case, the failure of the `G' property in this case has a very different character from that in \textbf{GR}, and it is much less obvious, in light of the \textbf{Ring} example, that the `G' property is really capturing what is wrong with the category \textbf{GR}, as a representation of general relativity.

Stepping back from this discussion, one might reasonably ask: if the `G' property is neither necessary nor sufficient for a category to have the features that we are interested in, then who cares?  I think the considerations just raised show that the `G' property is not quite what we want.  But this does not change the fact that some categories, such as \textbf{Set}, \textbf{Group}, and even \textbf{Ring}, seem to have some feature that \textbf{GR} appears to lack, regarding the way in which the network of arrows reflects or expresses the structure of the objects.  And this leads to a number of questions: Is there a property that better captures the intuition that motivated the `G' property?  Do all ``natural'' ``concrete'' categories (such as \textbf{Man}) share these features?  Does \emph{any} physical theory's category of models have this property?

\section{Where do we go from here?}\label{sec:roots}

In the paper thus far, I have proposed a certain informal sense in which some categories might be said to adequately represent the structure of a theory, and argued that the categories one encounters in many discussions of categorical equivalence in the philosophy of science literature do not seem to have the necessary features.  I have also proposed a formal condition intended to make the reasoning just sketched precise, but I have argued that this condition cannot be what we want.  This leads to a rather unsatisfactory situation: it seems that there is a sense in which the criterion of categorical equivalence, as discussed in the philosophical literature, is inadequate as a criterion of theoretical equivalence; and yet, it is not clear how the criterion should be modified, nor even how to precisely express how it fails.

In the present section, I will sketch three ways of responding to this situation.  Some philosophers have already begun exploring each of these three options.  But it seems to me that these options reflect importantly different conceptions of what it means to associate a category with a theory.  Thus, I think that the merits of each need to be carefully weighed in future work on this subject.

The first possible way forward would be to pursue the program suggested by section \ref{sec:Geroch}: we could find a `G'-like property that distinguishes \textbf{Set} from \textbf{GR} in a salient way, but which also respects the sort of symmetry exhibited by \textbf{Ring}.  We would then say that categorical equivalence yields theoretical equivalence only for theories whose associated categories satisfy the `G'-like property.  The main advantage of this approach is that it would capture what makes \textbf{Set} distinctive, and it would help directly diagnose the problem with \textbf{GR}.  In a sense, it is the brute-force solution to the problem.

On the other hand, this approach has a number of unattractive features.  Perhaps the most immediate is simply this: what could the property be?  One reason to be skeptical that any such property exists is that the idea the property is meant to express---that a category adequately represents the structure of a theory---is not obviously an intrinsic property of a category at all.  Instead, the feature we wish to capture concerns a relation that we wish to see between a category and a theory.  We want to know not just about the category itself, but whether it has certain capacities relative to some theory or other.  Of course, this is just a rough implausibility argument; it could well turn out that, precisely because categories with certain properties \emph{do} capture the structure of some theory, their capacity to do so is manifest, as it were, internally.

But let us suppose this strategy were successful, in the sense that some property could be found.  There are still reasons to doubt that it is the right way forward for the program.  First, we already have good reason to expect that this approach would limit the applicability of categorical equivalence, since the example of \textbf{GR} already suggests that categories that we might be interested in using to represent physical theories are unlikely to have the requisite property.  Indeed, it is not clear that we should expect \emph{any} physical theory, encountered in the wild, to naturally be associated with a category satisfying the sort of property envisaged by this program.

If this is right, then it would suggest that some or all of the theories for which categorical equivalence has already been used would no longer count as categorically (or theoretically) equivalent.  Fair enough, one might say: that is progress in our understanding of the relationship between these theories.  But if this is right, then the apparent successes of the categorical equivalence program become a mystery: as I noted above, it seems to me that categorical equivalence has given the correct verdicts in the cases to which it has been applied.  If categorical equivalence, as a criterion, is restricted only to theories whose categories have a certain property, and none of the theories considered thus far have that property, then we need to start from scratch in understanding in what sense the theories are equivalent.

So much for the first option, which seems to me to fail even if it succeeds.\footnote{That said: there remains an interesting question raised by this first approach, which is: if categories $C$ and $D$ are equivalent, with $F:C\rightarrow D$ realizing that equivalence, then what, if any, structural relationship holds between objects $c$ in $C$ and $F(c)$ in $D$?  I am grateful to Thomas Barrett for emphasizing this point, which I completely endorse.}  This leads to a second option, which would be to change the criterion of equivalence not by limiting attention to categories with certain properties, but by restricting attention to \emph{functors} with certain properties.  In particular, this is the sort of proposal that \citet{Hudetz} has defended: recall that on Hudetz's proposal, one should require that two theories are equivalent just in case they are categorically equivalent, where the functor realizing that equivalence is definable.\footnote{In a recent talk, Thomas Barrett described a similar, but distinct, program, on which it is ``well-behaved'' functors that realize equivalences.  I will not attempt to reconstruct (or scoop) his ideas here, but note only that it is another proposal that falls into this second category---or, perhaps, somewhere in between the first and second approaches, depending on how it is spelled out.}  This leads to a criterion of equivalence that Hudetz calls \emph{definable categorical equivalence}.

This approach has some virtues.  From this point of view, the problems described in previous sections arise not so much because some categories fail to capture the structure of theories, but rather because we compare those categories using a poorly behaved criterion.\footnote{One might be tempted by a possible resonance with the previous proposal, and try to modify the `G' property, using the notion of definable functor, as follows: a category satisfies the `H' property if every full, faithful, and essentially surjective \emph{definable} functor $F:C\rightarrow C$ is naturally isomorphic to $1_C$.  But this proposal is unlikely to work, since the functor $Op:$\textbf{Ring}$\rightarrow$\textbf{Ring} apparently counts as definable.} One might then conjecture that if we limited attention only to definable functors, the ``problematic'' examples of functors that realized (auto)equivalences in examples such as \textbf{GR} would go away, and we would be left only with examples such as \textbf{Ring}, where the failure of the `G' property did not seem to rule out the possibility that the category captured the internal structure of the objects of the category.

But attractive as this proposal may seem, we should be careful about what, exactly, it amounts to.  Recall the ideology above: I argued that to understand the ``structure'' of a mathematical gadget, we must study the maps that we take to preserve that structure.  In general, by changing the ``structure preserving'' maps we consider, we are implicitly changing the structures preserved by those maps.  And this is precisely what we are doing when we move from functors to definable functors as maps between categories.

Presumably something like this can be done, but we need to be careful and explicit.  In particular, no one has clearly articulated what sorts of mathematical gadgets are related by definable functors.\footnote{I do not mean to criticize Hudetz here.  He is explicit about the assumptions he is making when he defines definable functors, and makes clear why in the cases of interest, definable functors are well-defined.  But I read his assumptions as sufficient conditions for making sense of definable functors, which is weaker than a theory of the sorts of structures that definable functors relate.}  And this is a concern not only for conceptual reasons, but also for technical ones. For a functor $F:C\rightarrow D$ to be definable, certain properties must hold concerning languages associated with the objects of $C$ and $D$.  But a generic category does not have a language associated with its objects.\footnote{Observe that for some categories---toposes---there is an ``internal language'' associated with the category.  But this notion of internal language is not the same as the notion of ``language of objects of a category'' associated with definable functors.}  So in general, how can we evaluate whether a functor is definable?  The situation is strongly analogous to noting that not all functions between sets are well-behaved, and then restricting attention to continuous or smooth functions---without first defining a notion of ``topological space'' or ``manifold''.  To properly define a notion of definable functor, we first need to introduce a new kind of structure, a Hudetz category, where Hudetz category structure is whatever is preserved by definable functors.  

These remarks are not meant to dismiss the proposal.  To the contrary, I think it is a fruitful one to pursue.  But the remarks do suggest that the proposal is incomplete, and they raise questions about how much help the definable functor program will be.  In particular, as I have argued above, theories in the wild can, arguably, be associated with categories.  But it is much less clear that they can be associated with Hudetz categories, since whatever else is the case, it seems Hudetz categories will require one to specify a (possibly higher order) language associated with a physical theory, and it is not clear that there is a canonical choice of such a language.\footnote{Hudetz has recently made some progress in this direction: see, for instance, \citep{HudetzSignature}.} But suppose that this problem can be surmounted.  Then, even if suppose we \emph{can} associate a Hudetz category with any physical theory in a natural way, it seems the work will be done in identifying and justifying the choice of a language and establishing the necessary definability properties, which raises the concern that category theory plays little role. 

On the other hand, it is possible that although the proposal is still incomplete, it is unproblematic, for independent reasons.  Consider, again, the following analogy: suppose a ``definable functor'' is a bit like a ``continuous function''.  Of course, we need a topology to make sense of continuous functions.  But some spaces, such as $\mathbb{R}$, come equipped with a canonical topology, or a unique topology compatible with other structure.  And some functions of interest on $\mathbb{R}$, such as polynomials, are all automatically continuous in that topology. One might hope or expect that, although we have not yet defined the structure analogous to ``topology'' on categories of interest, once we do so we will find that there was a unique or canonical choice, and that the functors that seemed to be the ones of interest will automatically count as ``definable'' or otherwise well-behaved.  Indeed, it might be that we want to generate our definitions so that this turns out to be the case.  This possibility strikes me as the most optimistic for the program, though its status remains unclear.

Finally, I will now turn to the third option for a path forward.  To begin, consider again the examples of ``successes'' mentioned above---Einstein algebras and general relativity; Lagrangian and Hamiltonian mechanics; and so on.  Investigating the proofs of categorical equivalence in each of these cases, we find that the crucial step relies on some background, often deep, mathematical fact.  For instance, the relationship between vector potential and electromagnetic field formulations of electromagnetism ultimately comes down to Poincar\'e's lemma.  The relationship between ``standard'' and geometrized Newtonian gravitation ultimately depends on Trautman's theorems.  The relationship between Lagrangian and Hamiltonian mechanics depends on the Legendre transformation, and that between Einstein algebras and general relativity is s special case of function-space duality.  And so on.

But if it is really these relationships that are in the background, what is added by proving that there is a categorical equivalence (or inequivalence)?  The answer is that, in establishing a categorical equivalence, we show that these relationships are functorial, and then determine whether those functors are full, faithful, and essentially surjective.  In other words, one attempts to show that the mappings on objects determined by the relationships in question take every model of each theory to an essentially unique model of the other theory; and that it does so in such a way that every structure-preserving map between the models of one theory correspond uniquely to a structure-preserving map between the corresponding models of the other theory, and vice versa.  These are natural things to (try to) establish about any mathematical relationship, and establishing whether they hold in a particular case can underwrite a claim that a given relationship really does capture a sense in which two theories are equivalent.

Abstracting, then, from this discussion, one might say that what we are really doing when we establish that physical theories are categorically equivalent is abstracting ``pure category'' structure from a richer characterization of theories, and using that category structure to provide a heuristic for evaluating relationships of prior interest.  (I call this idea ``Rosenstock's heuristic'', because I think this perspective is adopted in much of Sarita Rosenstock's work on categorical equivalence \citep{RosenstockTDA}.)

One way of thinking about this proposal would be to say that categorical equivalence is necessary for equivalence of theories, but that it may not be sufficient.  Sufficiency, meanwhile, requires a more subtle and contextual analysis of the proposed relationships between theories---one that, in practice, often involves establishing that the relationships under consideration are already known to preserve ``empirical content'' in some substantive, but context-dependent, way.  In other words, much of the work is done by showing that there are alternative formulations of a theory that are, in some suitable sense, empirically equivalent; establishing that the relationship realizing that empirical equivalence is \emph{also} a categorical equivalence, then, provides a still stronger sense in which the theories should be said to be equivalent.\footnote{Compare this perspective to classic arguments due to \citet{Sklar}, recently amplified by, for instance, \citet{Coffey}, \citet{NguyenTE}, and \citet{Butterfield}, to the effect that a ``purely formal'' criterion of equivalence could never be adequate.  (Recall note \ref{fn:IE}.)  Here it is a semantic relationship---that is, a relationship between the interpreted, applied theories---that is ultimately the starting point, and then the formal methods are a guide to evaluating such relationships.  Consider, too, a connection to \citep{NortonUnderdetermination}, which argues that empirically equivalent theories are more or less certain to be equivalent in some stronger sense, or else to differ in ways that make more clearly preferable; from the present perspective, categorical (in)equivalence is a way of establishing how much, if anything, is missing from some empirical equivalence.}  And if the functor is \emph{not} an equivalence, then one can use the heuristic to better understand what is ``lost'' as one moves from one formulation to the other.\footnote{On this point, see the discussions in \citet{Baez+etal}, \citet{WeatherallUG}, \citet{Nguyen+etal}, and \citet{Bradley+etal}.}

From this perspective, the examples of \textbf{GR} and \textbf{Ring} have little bearing on the equivalence relationships under consideration: categorical equivalence is, in a sense, secondary---something we seek to establish only after determining that two theories are empirically equivalent.  The fact that there exist apparently pathological autoequivalences of \textbf{GR} is irrelevant because those autoequivalences are pathological precisely because they do not preserve empirical content.  They simply do not realize the sort of relationships that we are interested in.  And whether there are autoequivalences of \textbf{Ring} is of interest only if it turns out that \textbf{Ring} is associated with some physical theory, and those autoequivalences preserves empirical content.  Without that, they, too, are irrelevant.  Similarly, understanding categorical equivalence in terms of the Rosenstock heuristic explains why the ``successes'' noted above were, in fact, successes: they were all cases in which there existed a salient relationship between theories, the status of which was clarified by recasting it in categorical terms.

Still, there are disadvantages to adopting this perspective that are important to recognize.  Perhaps the most significant disadvantage is that on this view, much (but not all) of the work in establishing that two theories are theoretically equivalent falls back on the murky question of whether those theories are \emph{empirically} equivalent, which arguably makes the criterion of equivalence vaguer than it at first appears.  (On the other hand, insofar as empirical equivalence is necessary for theoretical equivalence, all of the approaches discussed in this section face this worry.)  A related concern is that, on the other approaches discussed, categorical equivalence is meant to capture some precise sense in which the mathematical structures used by two theories are equivalent, qua mathematical structures; empirical equivalence, then, establishes merely that in addition to being equivalent qua mathematics, the structures are used in compatible ways for representational purposes.  On the present perspective, this relationship is reversed.  The claim that categorical equivalence, or some modification thereof, should be expected to capture some robust notion of mathematical equivalence is dispensed with, which makes the significance of categorical equivalence more obscure.

This last set of remarks point a significant difference between the Rosenstock heuristic and both the `G' property and Hudetz category approaches.  On the first two approaches, a category of models, satisfying some further properties or endowed with some further structure, is intended to capture or represent a theory, full stop.  Implicit, I think, in these approaches is the idea that a theory---or at least, a mathematical theory, though perhaps also a physical theory---is the sort of thing that admits of some adequate, precise characterization, once and for all.\footnote{Defenders of these approaches might well balk at this point.  Do they really need to be committed to the view that categories of models are representations of theories ``once and for all''?  But if the goal is to determine if two theories are equivalent \emph{as theories}, then presumably that condition needs to capture everything salient about the pairs of theories.  If the goal is to offer a weaker notion of equivalence then much more needs to be said about what features the standard establishes equivalence with regards to.  This is what empirical equivalence offered: equivalence with regard to the predictions made by two theories, without implying ``full'' equivalence.  One possible line, here, would be to say that categorical equivalence and its various elaborations are attempting to capture a kind of ``structural equivalence'', though I think more needs to be said about just what that means.}  To pursue these approaches would be to evaluate whether various candidate representations of a theory succeed.  But the third approach I have discussed simply sets this idea aside.  One does not need to suppose that a theory is or can be represented by a given category in general; one merely needs to assume that for the purposes of evaluating certain features of a proposed relationship between theories, it is valuable to represent a theory by a category.

\section{Conclusion}

My goal in this paper has been to critically re-evaluate categorical equivalence as a criterion of theoretical equivalence for physical theories.  My worries turn on a prior question, of whether a category of models can be said to adequately capture the structure of a (mathematical) theory.  I have argued that the answer to this question is ``no'', at least in the general case, which then leads to a number of further questions.  One such question is whether one can express, as a precise condition on a category, a necessary or sufficient condition for that category to encode the internal structure of its objects.  Another question is where we should go from here, supposing that one accepts my arguments.  On this latter question, I offer three possible paths, each of which, I think, is suggested by work already in the literature, and I discuss some advantages and disadvantages of each.

I will conclude with two remarks.  One is just to clarify that the possible paths forward that I propose are not mutually exclusive, or, likely, exhaustive.  Indeed, I think all three \emph{should} be pursued, and that the fruits of each will bear on the others.\footnote{In particular, one might worry that the third approach brushes too many foundational questions aside, and that although it is pragmatically attractive, we should still be interested in the answers to those questions---which the first two approaches may yet yield.}  That said, as I have argued in the last section, these proposals seem to turn on different conceptions of what the purpose of introducing a category of models of a physical theory is meant to be, and so some care will be needed in future work to keep these different goals clearly in sight.  The second remark is that I wish to emphasize the tentative nature of the arguments here.  I am expressing worries---not proving theorems or even defending particular views.  From one perspective, this may make the paper seem unsatisfying or unclear.  I am sympathetic.  But I think the real issue, which I have tried to bring forward here, is that a program that has received considerable attention in recent years remains underspecified and inchoate, and it is my hope that the considerations raised here help move the project forward.

\section*{Acknowledgments}

I am grateful to Thomas Barrett, Lukas Barth, Neil Dewar, Ben Eva, Ben Feintzeig, Hans Halvorson, Laurenz Hudetz, David Malament, Toby Meadows, and Sarita Rosenstock for many helpful conversations in connection with this material, and to Hajnal Andr\'eka, Thomas Barrett, Istvan N\'emeti, and an anonymous referee for detailed comments on a previous draft.  A version of the paper was presented at a workshop at the Munich Center for Mathematical Philosophy; I am grateful to the organizers and the audience for their valuable feedback.

\bibliographystyle{elsarticle-harv}
\bibliography{equivalence}

\begin{thebibliography}{62}
\expandafter\ifx\csname natexlab\endcsname\relax\def\natexlab#1{#1}\fi
\expandafter\ifx\csname url\endcsname\relax
  \def\url#1{\texttt{#1}}\fi
\expandafter\ifx\csname urlprefix\endcsname\relax\def\urlprefix{URL }\fi

\bibitem[{Andr{\'e}ka and N{\'e}meti(2014)}]{Andreka+NemetiComparing}
Andr{\'e}ka, H., N{\'e}meti, I., 2014. Comparing theories: the dynamics of
  changing vocabulary. In: Johan van Benthem on logic and information dynamics.
  Springer, pp. 143--172.

\bibitem[{Andr\'eka and N\'emeti(2014)}]{Andreka+NemetiNotes}
Andr\'eka, H., N\'emeti, I., 2014. Definability theory course notes, available
  at http://www.math-inst.hu/pub/algebraic-logic/DefThNotes0828.pdf. Accessed
  July 2015.

\bibitem[{Awodey and Forssell(2013)}]{Awodey+Forssell}
Awodey, S., Forssell, H., 2013. First-order logical duality. Annals of Pure and
  Applied Logic 164~(3), 319--348.

\bibitem[{Baez et~al.(2004)Baez, Bartel, and Dolan}]{Baez+etal}
Baez, J., Bartel, T., Dolan, J., 2004. Property, structure, and stuff,
  available at: http://math.ucr.edu/home/baez/qg-spring2004/discussion.html.

\bibitem[{Baez and Schreiber(2007)}]{Baez+Schreiber}
Baez, J., Schreiber, U., 2007. Higher gauge theory. In: Davydov, A. (Ed.),
  Categories in Algebra, Geometry, and Mathematical Physics. American
  Mathematical Society, Providence, RI, pp. 7--30.

\bibitem[{Bain(2003)}]{Bain}
Bain, J., 2003. Einstein algebras and the hole argument. Philosophy of Science.

\bibitem[{Barrett(2014)}]{BarrettCM1}
Barrett, T., 2014. On the structure of classical mechanics. The British Journal
  for the Philosophy of Science 66~(4), 801--828.

\bibitem[{Barrett(2017)}]{BarrettCM2}
Barrett, T.~W., 2017. Equivalent and inequivalent formulations of classical
  mechanics. British Journal for Philosophy of ScienceForthcoming.
  http://philsci-archive.pitt.edu/13092/.

\bibitem[{Barrett and Halvorson(2016)}]{Barrett+HalvorsonME}
Barrett, T.~W., Halvorson, H., 2016. Morita equivalence. The Review of Symbolic
  Logic 9~(3), 556--582.

\bibitem[{Barth(2018)}]{Barth}
Barth, L., 2018. Tk. Master's thesis, University of Heidelberg.

\bibitem[{Belot(1998)}]{BelotUE}
Belot, G., 1998. Understanding electromagnetism. The British Journal for the
  Philosophy of Science 49~(4), 531--555.

\bibitem[{Benacerraf(1965)}]{Benacerraf}
Benacerraf, P., 1965. What numbers could not be. Philosophical Review 74~(1),
  47--73.

\bibitem[{Bradley(2019)}]{Bradley}
Bradley, C., 2019. The non-equivalence of einstein and lorentz. The British
  Journal for the Philosophy of Science. Forthcoming. DOI:10.1093/bjps/axz014.

\bibitem[{Bradley and Weatherall(2020)}]{Bradley+etal}
Bradley, C., Weatherall, J.~O., 2020. On representational redundancy, surplus
  structure, and the hole argument. Foundations of PhysicsForthcoming.
  arXiv:1904.04439 [physics.hist-ph].

\bibitem[{Brighouse(Forthcoming)}]{BrighouseNew}
Brighouse, C., Forthcoming. Confessions of a (cheap) sophisticated
  substantivalist. Foundations of Physics.

\bibitem[{Brunetti et~al.(2003)Brunetti, Fredenhagen, and
  Verch}]{Brunetti+etal}
Brunetti, R., Fredenhagen, K., Verch, R., 2003. The generally covariant
  locality principle--a new paradigm for local quantum field theory.
  Communications in Mathematical Physics 237~(1-2), 31--68.

\bibitem[{Burgess(2015)}]{Burgess}
Burgess, J.~P., 2015. Rigor and Structure. Oxford University Press, New York.

\bibitem[{Butterfield(2019)}]{Butterfield}
Butterfield, J., 2019. On dualities and equivalences between physical theories.
  In: Huggett, N., W\"uthrich, C. (Eds.), Spacetime After Quantum Gravity.
  Forthcoming.

\bibitem[{Coffey(2014)}]{Coffey}
Coffey, K., 2014. Theoretical equivalence as interpretive equivalence,
  forthcoming from \emph{The British Journal for the Philosophy of Science}.

\bibitem[{Curiel(2013)}]{Curiel}
Curiel, E., 2013. Classical mechanics is {L}agrangian; it is not {H}amiltonian.
  The British Journal for Philosophy of Science 65~(2), 269--321.

\bibitem[{Dewar and Eva(2017)}]{Dewar+Eva}
Dewar, N., Eva, B., 2017. A categorical perspective on symmetry and
  equivalence.

\bibitem[{Earman(1986)}]{Earman1986}
Earman, J., 1986. Why space is not a substance (at least not to first degree).
  Pacific Philosophical Quarterly 67~(4), 225--244.

\bibitem[{Earman(1989)}]{EarmanWEST}
Earman, J., 1989. World Enough and Space-Time. The MIT Press, Boston.

\bibitem[{Earman and Norton(1987)}]{Earman+Norton}
Earman, J., Norton, J., 1987. What price spacetime substantivalism? {T}he hole
  story. The British Journal for the Philosophy of Science 38~(4), 515--525.

\bibitem[{Formica and Friend(2020)}]{FormicaFriend}
Formica, G., Friend, M., 2020. In the footsteps of hilbert: The
  andr\'eka-n\'emeti group's logical foundations of theories in physics. In:
  Madarasz, J., Szekely, G. (Eds.), Hajnal Andréka and István Németi on
  unity of science: from computing to relativity theory through algebraic
  logic. Springer, Heidelberg, . This volume.

\bibitem[{Freyd(1964)}]{Freyd}
Freyd, P.~J., 1964. Abelian categories. Vol. 1964. Harper \& Row New York.

\bibitem[{Geroch(1972)}]{GerochEA}
Geroch, R., 1972. Einstein algebras. Communications in Mathematical Physics 26,
  271--275.

\bibitem[{Halvorson(2012)}]{Halvorson}
Halvorson, H., 2012. What scientific theories could not be. Philosophy of
  Science 79~(2), 183--206.

\bibitem[{Hudetz(2018)}]{Hudetz}
Hudetz, L., 2018. Definable categorical equivalence. Philosophy of
  ScienceForthcoming. http://philsci-archive.pitt.edu/14297/.

\bibitem[{Hudetz(2019)}]{HudetzSignature}
Hudetz, L., 2019. The semantic view of theories and higher-order languages.
  Synthese 196~(3), 1131--1149.

\bibitem[{Jacobson(1951)}]{JacobsonAlgebra}
Jacobson, N., 1951. Lectures in Abstract Algebra, Vol. 1: Basic Concepts, The
  University Series in Higher Mathematics. D. Van Nostrand Co., Inc.,
  Princeton, NJ.

\bibitem[{Lam(2013)}]{Lam}
Lam, T.-Y., 2013. A first course in noncommutative rings. Vol. 131. Springer
  Science \& Business Media, Heidelberg.

\bibitem[{Lawvere(1964)}]{Lawvere}
Lawvere, F.~W., 1964. An elementary theory of the category of sets. Proceedings
  of the national academy of sciences 52~(6), 1506--1511.

\bibitem[{Lefever and Sz{\'e}kely(2018)}]{Lefever+Szekely}
Lefever, K., Sz{\'e}kely, G., 2018. On generalization of definitional
  equivalence to languages with non-disjoint signatures, arXiv:1802.06844.

\bibitem[{Leinster(2014)}]{Leinster}
Leinster, T., 2014. Basic Category Theory. Cambridge University Press,
  Cambridge.

\bibitem[{Lurie(2018)}]{Lurie}
Lurie, J., 2018. Ultracategories,
  http://www.math.harvard.edu/~lurie/papers/Conceptual.pdf.

\bibitem[{{Mac Lane}(1998)}]{MacLane}
{Mac Lane}, S., 1998. Categories for the Working Mathematician, 2nd Edition.
  Springer, New York.

\bibitem[{Makkai(1993)}]{Makkai}
Makkai, M., 1993. Duality and definability in first order logic. American
  Mathematical Soc., Providence, RI.

\bibitem[{Manchak(2020)}]{ManchakAN}
Manchak, J., 2020. General relativity as a collection of collections of models.
  In: Madarasz, J., Szekely, G. (Eds.), Hajnal Andréka and István Németi on
  unity of science: from computing to relativity theory through algebraic
  logic. Springer, Heidelberg, . This volume.

\bibitem[{Nestruev(2003)}]{Nestruev}
Nestruev, J., 2003. Smooth manifolds and observables. Springer, Berlin.

\bibitem[{Nguyen(2017)}]{NguyenTE}
Nguyen, J., 2017. Scientific representation and theoretical equivalence.
  Philosophy of Science 84~(5), 982--995.

\bibitem[{Nguyen et~al.(2018)Nguyen, Teh, and Wells}]{Nguyen+etal}
Nguyen, J., Teh, N.~J., Wells, L., 2018. Why surplus structure is not
  superfluous. British Journal for Philosophy of ScienceForthcoming.

\bibitem[{North(2009)}]{North}
North, J., 2009. The `structure' of physics: A case study. Journal of
  Philosophy 106~(2), 57--88.

\bibitem[{Norton(2008)}]{NortonUnderdetermination}
Norton, J., 2008. Must evidence underdetermine theory. In: Kourany, J.~A.,
  Carrier, M., Howard, D. (Eds.), The challenge of the social and the pressure
  of practice: Science and values revisited. University of Pittsburgh Press
  Pittsburgh, Pittsburgh, PA, pp. 17--44.

\bibitem[{Norton(2011)}]{NortonSEP}
Norton, J.~D., 2011. The hole argument. In: Zalta, E.~N. (Ed.), The Stanford
  Encyclopedia of Philosophy, {F}all 2011 Edition.
  {http://plato.stanford.edu/archives/fall2011/entries/spacetime-holearg/}.

\bibitem[{Rosenstock(2019)}]{RosenstockTDA}
Rosenstock, S., 2019. A categorical consideration of physical formalisms. Ph.D.
  thesis, University of California, Irvine.

\bibitem[{Rosenstock et~al.(2015)Rosenstock, Barrett, and
  Weatherall}]{Rosenstock+etal}
Rosenstock, S., Barrett, T.~W., Weatherall, J.~O., 2015. On {E}instein algebras
  and relativistic spacetimes. Studies in History and Philosophy of Science
  Part B: Studies in History and Philosophy of Modern Physics 52, 309--316.

\bibitem[{Rosenstock and Weatherall(2016)}]{Rosenstock+Weatherall}
Rosenstock, S., Weatherall, J.~O., 2016. A categorical equivalence between
  generalized holonomy maps on a connected manifold and principal connections
  on bundles over that manifold. Journal of Mathematical Physics 57~(10),
  102902.

\bibitem[{Rynasiewicz(1992)}]{Ryno}
Rynasiewicz, R., 1992. Rings, holes and substantivalism: On the program of
  leibniz algebras. Philosophy of Science 59~(4), 572--589.

\bibitem[{Sklar(1982)}]{Sklar}
Sklar, L., 1982. Saving the noumena. Philosophical Topics 13~(1), 89--110.

\bibitem[{van Fraassen(1980)}]{VanFraassenSI}
van Fraassen, B., 1980. The Scientific Image. Oxford University Press, Oxford.

\bibitem[{van Fraassen(2008)}]{vanFraassenSR}
van Fraassen, B., 2008. Scientific Representation. Oxford University Press,
  Oxford.

\bibitem[{van Oosten(2002)}]{vanOosten}
van Oosten, J., 2002. Basic Category Theory. BRICS Lecture Series LS-95-01,
  http://www.staff.science.uu.nl/∼ooste110/www/syllabi/catsmoeder.pdf.

\bibitem[{Weatherall(2016{\natexlab{a}})}]{WeatherallTE}
Weatherall, J.~O., 2016{\natexlab{a}}. Are {N}ewtonian gravitation and
  geometrized {N}ewtonian gravitation theoretically equivalent? Erkenntnis
  81~(5), 1073--1091.

\bibitem[{Weatherall(2016{\natexlab{b}})}]{WeatherallYM}
Weatherall, J.~O., 2016{\natexlab{b}}. Fiber bundles, yang--mills theory, and
  general relativity. Synthese 193~(8), 2389--2425.

\bibitem[{Weatherall(2016{\natexlab{c}})}]{WeatherallHoleArg}
Weatherall, J.~O., 2016{\natexlab{c}}. Regarding the ‘hole argument’. The
  British Journal for the Philosophy of Science 69~(2), 329--350.

\bibitem[{Weatherall(2016{\natexlab{d}})}]{WeatherallUG}
Weatherall, J.~O., 2016{\natexlab{d}}. Understanding gauge. Philosophy of
  Science 83~(5), 1039--1049.

\bibitem[{Weatherall(2017)}]{WeatherallLandry}
Weatherall, J.~O., 2017. Category theory and the foundations of classical
  space-time theories. In: Landry, E. (Ed.), Categories for the Working
  Philosopher. Oxford University Press, Oxford, pp. 329--348.

\bibitem[{Weatherall(2019{\natexlab{a}})}]{WeatherallReview1}
Weatherall, J.~O., 2019{\natexlab{a}}. Theoretical equivalence in physics, part
  1. Philosophy Compass 14~(5), e12592.

\bibitem[{Weatherall(2019{\natexlab{b}})}]{WeatherallReview2}
Weatherall, J.~O., 2019{\natexlab{b}}. Theoretical equivalence in physics, part
  2. Philosophy Compass 14~(5), e12591.

\bibitem[{Weatherall(2020)}]{WeatherallStein}
Weatherall, J.~O., 2020. Some philosophical prehistory of the (earman-norton)
  hole argument. Studies in History and Philosophy of Modern
  PhysicsForthcoming. arXiv:1812.04574 [physics.hist-ph].

\bibitem[{Winnie(1986)}]{Winnie}
Winnie, J.~A., 1986. Invariants and objectivity: A theory with applications to
  relativity and geometry. In: Colodny, R. (Ed.), From Quarks to Quasars.
  University of Pittsburgh Press, Pittsburgh, pp. 71--180.

\end{thebibliography}

\end{document}